# Motion of charged particles in bright squeezed vacuum

**Matan Even Tzur and Oren Cohen**

Solid State Institute and Physics Department, Technion-Israel Institute of Technology, Haifa 3200003, Israel.

**Abstract**

**The motion of laser-driven electrons quivers with an average energy termed pondermotive energy. We explore electron dynamics driven by bright squeezed vacuum (BSV), finding that BSV induces width oscillations, akin to electron quivering in laser light, with an equivalent ponderomotive energy. In the case of bound electrons, width oscillations may lead to tunnel ionization with noisy sub-cycle structure. Our results are foundational for strong-field and free-electron quantum optics, as they shed light on tunnel ionization, high harmonic generation, and nonlinear Compton scattering in BSV.**

The ponderomotive energy scale $U_p$ is the cycle averaged energy of a charged particle interacting with a classical, monochromatic, linearly polarized, electric field. It is a key figure of merit in the theory of high-field ionization [1,2], high harmonic generation [3,4], and plasma physics [5,6]. Compared to the natural energy scales of a system at hand, the scale of $U_p$ determines the transition between regimes in light matter interactions. A prominent example is the transition between the multi-photon and tunnel ionization regimes of atoms, which is determined by the Keldysh parameter [1] $\gamma = \sqrt{I_p/2U_p}$, in which $I_p$ is the ionization potential of an atom. $U_p$ is given by the famous formula (atomic units throughout):

$$U_p^{(c)} = e^2 E_a^2 \Big/ 4m\omega_p^2 \qquad (1)$$

in which $e$ is the electron charge, $E_a$ is the peak amplitude of a monochromatic & linearly polarized electric field of frequency $\omega_p$, and $m$ is the mass of the electron. A superscript $(c)$ was added to indicate that this ponderomotive energy is 'classical', in the sense that it corresponds to the energy of a classical electron in a classical field, as derived by Newton's equations of motion.

The ponderomotive energy scale is associated with *quiver motion*. A quivering particle oscillates back and forth in space at the frequency of a driving laser, following a trajectory prescribed by the driving laser field. While almost interchangeable with light-induced motion, any sinusoidally oscillating force will induce a quiver motion with an average ponderomotive energy. It happens that in the case of (classical) electromagnetic waves & charged particles, this force is a coherent electric field. Indeed, if light carries a vanishing coherent electric field amplitude $E_a = 0$, the sinusoidally oscillating force vanishes, and Eq. (1) yields $U_p^{(c)} = 0$ (i.e., the particle stands still).

At the same time, there are many indications that coherent motion of matter may be induced even by quantum fields carrying a vanishing electric field amplitude. A recent prominent example in this context is the interaction of bright squeezed vacuum (BSV) with various phases of matter, in different regimes [7]. Perturbative nonlinear optical processes [8,9] and photoionization [10] driven by BSV were already observed experimentally, while non-perturbative nonlinear optics [11,12] and nonlinear Compton scattering [13] driven by BSV are predicted to host a variety of novel phenomena. These studies are all indicative that BSV induces electronic motion in the medium, even though it carries electric field *fluctuations,* yet a vanishing coherent electric field amplitude. Therefore, quiver motion &

ponderomotive energy, as they are currently formulated and conceptualized, are invalid for these interactions.

Here we generalize the concepts of quiver motion and its associated ponderomotive energy for charged particles in quantum light fields, focusing on the multi-mode squeezed vacuum state. **More generally, we explore *sub-cycle motion* of free & bound particles driven by bright squeezed vacuum.** We begin by calculating numerically the motion of a free electron driven by a bright squeezed vacuum field. We find that free electrons in bright squeezed vacuum undergo width oscillations, i.e., coherent stretching and squeezing in real space. We show numerically and analytically that average energy associated with these width oscillations, namely, the quantum ponderomotive energy $U_p^{(q)}$, is equal to the classical ponderomotive energy $U_p^{(c)}$. In the case of bound electrons, we find that such width oscillations may be violent enough to induce tunnel ionization (and recombination), which follow noisy sub-cycle dynamics. Our results are foundational for extreme nonlinear quantum optics, as they provide insight to the underlying mechanisms of nonlinear Compton scattering [13], HHG [11,12], and tunnel ionization [10] when they are driven by squeezed vacuum.

**Free electron width oscillations and their ponderomotive energy**

We begin by calculating numerically the motion of a free electron placed in a bright-squeezed vacuum field. We performed three ab-initio time evolution calculations for an electron that initially occupies a gaussian wave-packet in 1D real space $|g\rangle \propto \exp(-x/4\sigma_0^2)$ and interacts with: **(i)** a single mode of EM vacuum at frequency $\Omega$, $|0_\Omega\rangle$, **(ii)** a coherent state $\widehat{D}(\alpha)|0_\Omega\rangle$, and **(iii)** a single mode of squeezed vacuum $\hat{S}(r)|0_\Omega\rangle$. Here, $\widehat{D}(\alpha)$ and $\hat{S}(r)$ are coherent shift and squeezing operators for the temporal mode $\Omega$, respectively [14]. Time evolution of the initial light-matter state under the Hamiltonian $\widehat{H} = \hat{p}^2/2m + \hat{x} \cdot \hat{E}_\Omega(t)$ is implemented through the $(t, t')$ method [15] (SI section II). Here, $\hat{x}$ and $\hat{p}$ are the electron position and momentum operators respectively, and $\hat{E}_\Omega(t)$ is the electric field operator of the $\Omega$ mode (pump). These calculations yield the time-dependent light matter state $|\Psi(t)\rangle$ for each initial light state. Finally, a partial trace on the photonic degrees of freedom is implemented, resulting in the reduced density matrix of the electron $\rho_{x,x'}^{(e)}(t)$, whose diagonal is the real-space wavefunction density $|\psi_e(x,t)|^2$. Figure 1. (a-c) present the wavefunction density of the electron coupled to EM vacuum and $|\psi_e(x,t)|^2$ for the three examined cases. Notably, while the coherent state induces a quivering displacement motion $\langle\hat{x}(t)\rangle_{CS} \neq 0$, that matches the Newtonian trajectory of a charged particle, the squeezed vacuum state results in a vanishing displacement $\langle \hat{x}(t) \rangle_{SV} = 0$. Examining the width of the wavepackets $\Delta X^2 = \langle \hat{x}^2(t) \rangle - \langle \hat{x}(t) \rangle^2$, we find that for the EM vacuum & coherent state fields, the Gaussian wavepacket naturally diffracts (expands) according to the analytical formula $\Delta X^2(t) = \sigma_0^2(1 + t^2/4\sigma_0^4)$ (Figure 1.(d)). In contrast, the width of the SV driven electron is periodically modulated, exhibiting coherent stretching & squeezing dynamics in real space, superimposed on the quadratic diffraction of the Gaussian wavepacket. Figure 1.(e) presents the time dependent kinetic energy of the electron obtained from the numerical calculation, $E_{\text{kin}}(t)$. The cycle-average of $E_{\text{kin}}(t)$ is the quantum-optical generalization of the ponderomotive energy, $U_p^{(q)} = \frac{1}{T} \int_0^T E_{\text{kin}}(t) dt$. The numerical calculation reveals that the $U_p^{(q)}$ is exactly equal to the classical ponderomotive energy imposed by an equally intense coherent state:

$$U_p^{(q)} = \frac{2e^2}{m\epsilon_0 c} \frac{I_{\text{vac}}}{4\Omega^2} \qquad (2)$$

Here, $m, e$ are the mass and charge of the electron, $\epsilon_0$ vacuum permittivity, $c$ speed of light. $I_{\text{vac}} \equiv c\hbar\Omega N_{SV}/V$ is the intensity of a squeezed vacuum beam with $N_{SV}$ photons in a quantization volume $V$, and a frequency $\Omega$. The number of photons is given by $N_{SV} = \sinh^2(r)$ where $r$ is the dimensionless squeezing parameter. In section I of the SI, we derive Eq. (2) analytically using 2$^{\text{nd}}$ order perturbation theory, and generalize it to multi-mode squeezed light, accounting for various forms of squeezed light such as polarization squeezed light and more. We find that the classical formula for ponderomotive energy, derived through Newton's equations of motion, applies to any form of multi-mode squeezed light. Furthermore, as a corollary of our calculation, we find the electron experiences a mass renormalization contingent upon vacuum squeezing; although miniscule for non-relativistic velocities (Eq. I.24 in the SI).

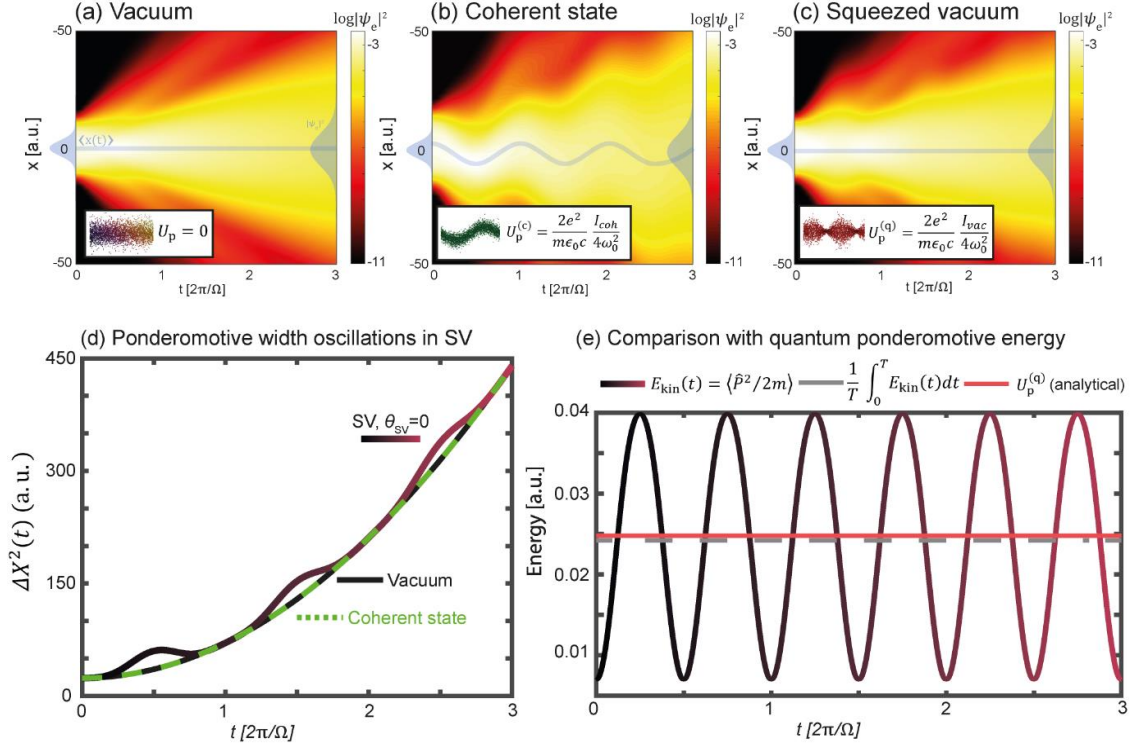

*Figure 1: **Time evolution of a free electron in quantum fields.** (a-c) Wavefunction density $|\psi_e(x,t)|^2$ for an electron coupled to (a) quantized electromagnetic vacuum (b) a coherent state (c) a single mode of squeezed vacuum. (d) Wave packet width as a function of time for the three cases plotted in (a-c). (e) time dependent expectation value of the kinetic energy. The time averaged energy (gray, dashed line) matches the analytically derived ponderomotive energy (pink, solid line).*

**Closed & open trajectories of the width**

When a charged particle is released in a classical electromagnetic field $\propto \cos(\omega t)$ at a position $x = 0$ and time $t = t_0$, it may exhibit either closed or open motion (Figure 2.(a)). If the electron started its motion after the peak (node) of the field, i.e., $0 < t_0/T < 0.25$ ($0.25 < t_0/T < 0.5$), it will (not) revisit $x = 0$ at a later time, resulting in a closed (open) trajectory. The notion of closed & open trajectories is central to the description HHG and ATI when driven by a coherent state. For instance, if electron starts its motion at the ionization time $t_0$, only ionization times in the range $0 < t_0/T < 0.25$ contribute to HHG and re-scattering ATI, because they result in closed trajectories. An analogous phenomenon exists in the context of the motion of charged particles in squeezed vacuum (Figure 2.(b)). A free electron that begins its motion after the anti-squeezed peak of the SV field variance will exhibit a closed trajectory in the sense that its width will revisit the free propagation width at a later time (at least once). Likewise, if the electron is released to motion right after the maximal squeezing of the field $0.25 < t_0/T < 0.5$ it will exhibit an open trajectory, i.e., it will always be wider than the free propagation width. To understand this motion, we consider the perturbative formula for the density matrix of the electron $\rho^{(e)}$, after interaction with the squeezed vacuum. The formula, derived using a quasi-probability distribution approach, reads [12]:

$$\rho^{(e)} \approx \frac{1}{\sqrt{2\pi}|E_{\text{vac}}|} \int dE_\alpha e^{-\frac{|E_\alpha|^2}{2|E_{vac}|^2}} |\phi_{E_\alpha}(t)\rangle\langle\phi_{E_\alpha}(t)| \tag{3}$$

$$I_{vac} \equiv \frac{1}{2}\epsilon_0 c |E_{vac}|^2 \tag{4}$$

where $I_{vac}$ is the intensity of the squeezed vacuum beam, and $|E_{vac}|$ is the electric field amplitude of an equally intense Glauber coherent state. It is also approximately the amplitude of electric field fluctuations in the anti-squeezed quadrature of the pump. The wavefunction $|\phi_{E_\alpha}(t)\rangle$ solves the following time dependent Schrodinger equation of a free electron in a classical electric field $E_\alpha = \langle\alpha|\hat{E}(t)|\alpha\rangle$

$$i\hbar \frac{\partial |\phi_{E_\alpha}(t)\rangle}{\partial t} = \left[-\frac{1}{2m}\nabla^2 - ez \cdot E_\alpha \cos(\Omega t)\right] |\phi_{E_\alpha}(t)\rangle \tag{5}$$

With the initial condition $|\phi_\alpha(t=0)\rangle = |g\rangle\langle g|$. According to equations (3)(5), the width of the wavepacket is given by

$$\Delta X^2(t) \approx \underbrace{\sigma_0^2 + \frac{t^2}{4\sigma_0^2}}_{\text{vacuum expansion}} + \underbrace{\langle x_{E_{\text{vac}}}(t)\rangle^2}_{\text{Width oscillations}} \tag{6}$$

Where $\langle x_{E_{\text{vac}}}(t)\rangle$ is the classical displacement of $|\phi_{E_\alpha}(t)\rangle$ with $E_\alpha = E_{\text{vac}}$, i.e., the displacement an equally intense coherent state would impose. Because $\langle x_{E_{\text{vac}}}(t)\rangle$ results in closed trajectories between $0 < t_0 < 0.25T$, so does $\Delta X^2(t)$. For $0.25T < t_0 < 0.5T$, the displacement $\langle x_{E_{\text{vac}}}(t)\rangle$ represents an open trajectory, therefore $\Delta X^2(t)$ does not touch the vacuum expansion line.

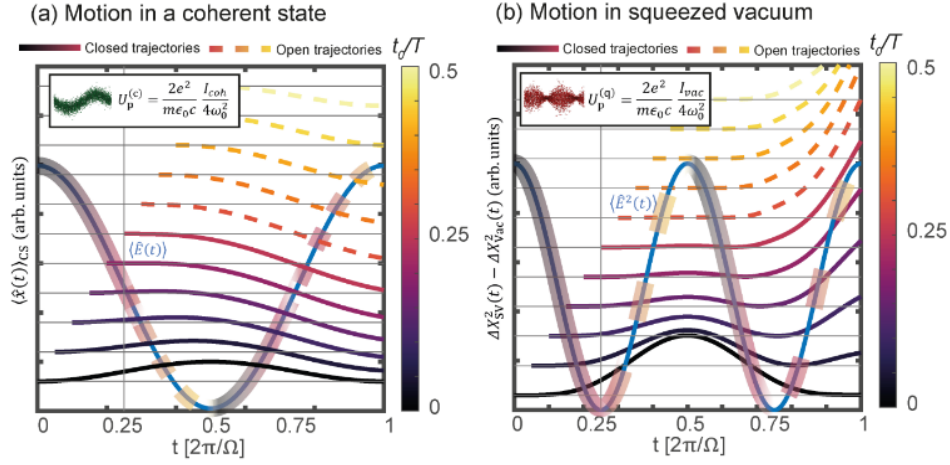

*Figure 2: **Closed and open trajectories of free electrons**. (a) The displacement of a free electron coupled to a coherent state for various initial moments $t_0$. The electric field is depicted in blue. (**b**) The excess width of a free electron driven by squeezed vacuum for various initial moments. Both observables (a-b) exhibits "closed" trajectories for initial times $t_0$ between 0 and 0.25T (solid lines), and open trajectories between 0.25T and 0.5T (dashed lines).*

Similarly to the case of a coherent state driver, we anticipate that only closed trajectories of the width contribute to HHG and re-scattering ATI, when they are driven by squeezed vacuum. Our reasoning is as follows: Eq.(9), that describes the width trajectories $\langle x^2(t) \rangle$ is calculated from a superposition of states. When the width trajectory is closed, the displacement of each individual state $\langle x_{E_\alpha}(t) \rangle$ in the superposition exhibits a closed trajectory, which leads to high harmonics & re-scattering ATI. Similarly, for open width trajectories, the individual constituents of the superposition do not follow closed displacement trajectories, and therefore do not contribute to HHG & re-scattering ATI.

**Sub-cycle dynamics of photoionization driven by bright-squeezed vacuum**

Next, we explore the dynamics of a bound electron in bright squeezed vacuum by adding a model Xe atomic potential to the numerical calculation [16]. This model atom supports two bound states with energies $E_g = 0.44\,a.u.$ and $E_e = -0.14\,a.u.$, as well as a third weakly bound state with an energy $\sim 0.00014\,a.u.$ and a continuum. We calculate the time evolution of the ground state of this atom $|g.s.\rangle_{Xe}$ when it is driven by BSV and coupled to one quantized radiation mode (SI section II). Again, we obtain the time-dependent light-matter state $\hat{\rho}(t)$ and perform a partial trace on the photonic degrees of freedom to obtain the reduced density matrix of the atom $\rho^{(\text{atom})}$. Figure 3. (b) presents the occupations of the ground and 1st excited states $\rho_{gg}^{(\text{atom})}$ and $\rho_{ee}^{(\text{atom})}$, showing Rabi-like oscillations between these levels. Their total occupation is nearly flat with a negative slope, with the slope being the time averaged tunnel ionization rate. Figure 3.(c). shows the time-dependent density of the electron in real space. It is evident that the electron undergoes tunnel ionization during the interaction, but unlike tunnel ionization in a coherent state of light, this tunnel ionization is symmetric to both sides of the atomic potential well. Figure 3.(d) shows the time derivative of the total occupation of all continuum states, indicative of the rate of tunnel ionization. We observe that this quantity exhibits rapid oscillations between positive and negative rates, i.e., the electron is rapidly switching from net ionization to net recombination. The amplitude of the oscillations is in phase with the amplitude of electric field fluctuations of the squeezed vacuum. We anticipate light emission (and particularly HHG) to occur during the recombination bursts, as the electron must release its kinetic energy in the form of radiation.

To summarize, we have generalized the classical notions of quiver motion and ponderomotive energy to a quantum optical context. We have shown theoretically that the motion of a free electron in squeezed vacuum consists of periodic stretching and squeezing of its wavefunction in real space and discovered that it exhibits open and close trajectories, in a similar fashion to displacement trajectories of electrons driven by coherent light. For the case of a bound electron, our results reveal the underlying dynamics associated with high harmonic generation driven by BSV and resolve sub-cycle features of tunnel ionization,. Additionally, we found the energy of an electron interacting with generalized multi-mode squeezed vacuum field to be exactly equal to the classical ponderomotive energy (derived by Newton's equations of motion). Our treatment holds for any mutli-mode form of squeezed light, for example, bi-chromatic two-mode squeezed vacuum [17], as well as for polarization squeezed light [18]. Looking forward, we expect our results to be directly applicable to nonlinear Compton scattering [13] driven by BSV and free electron shaping by quantum light [19,20], as both of these interactions are concerned with a free electron placed in a squeezed vacuum field. Additionally, as the presented results generalize the standard building blocks ($U_p$ and quiver motion) of strong-field physics to the quantum optical regime, we believe our results will play a foundational role in the emerging field of quantum-optical strong-field physics, which ranges from quantum information processing [21], explorations of light-matter entanglement [22–27], and more [28–32].

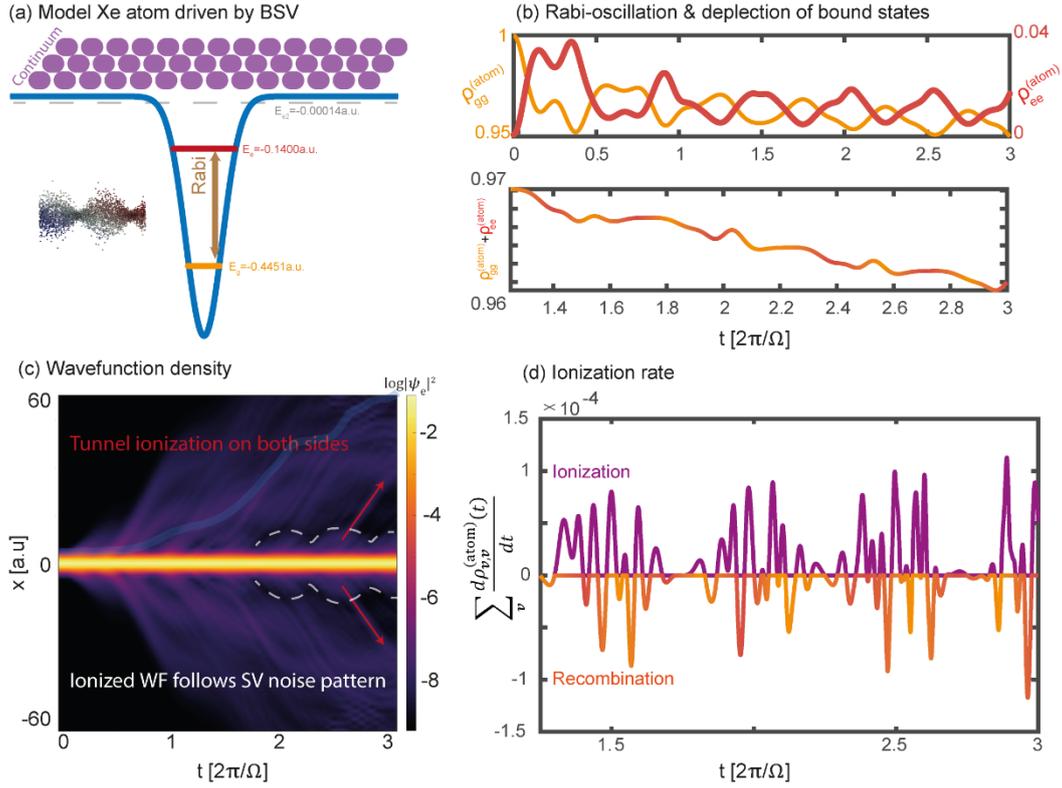

*Figure 3:* **Sub-cycle dynamics of bound electrons driven by BSV** *(a) illustration of a model Xe atom interacting with a squeezed vacuum field (b) populations of the ground and 1$^{st}$ excited state as a function of time. The states undergo Rabi oscillations, while steadily depleting as a function of time due to tunnel ionization. (c) Electron wavefunction density in real space as a function of time $|\psi_e(x,t)|^2$ (d) time derivation of total continuum state populations, indicative of the ionization rate as a function of time. The ionization occurs in noisy bursts that are in phase with the squeezed vacuum noise.*